\documentclass[final,1p,times]{elsarticle}

 \usepackage{graphics}
\usepackage{url}
\usepackage{amssymb}

\usepackage{color}

\begin{document}

\begin{frontmatter}

\title{Limited individual attention and online virality\\of low-quality information}

\author{
	Xiaoyan Qiu$^{1,2}$\footnote{These two authors contributed equally to this work}}
\author{Diego F. M. Oliveira$^{2*}$}
\author{Alireza Sahami Shirazi$^{3}$}
\author{Alessandro Flammini$^{2,4}$}
\author{Filippo Menczer$^{2,3,4}$}
\address{$^{1}$ School of Economics and Management - Shanghai Institute of Technology\\
	$^{2}$ Center for Complex Networks and Systems Research- School of Informatics and Computing - Indiana University\\
	$^{3}$ Yahoo Research\\
	$^{4}$ Indiana University Network Science Institute
}

\begin{abstract}
Social media are massive marketplaces where ideas and news compete for our attention~\cite{simon_designing_1971}. Previous studies have shown that  quality is not a necessary condition for online virality~\cite{refweng} and that knowledge about peer choices can distort the relationship between quality and popularity~\cite{Salganik854}. However, these results do not explain the viral spread of low-quality information, such as the digital misinformation that threatens our democracy~\cite{Forum:2013}. We investigate quality discrimination in a stylized model of online social network, where individual agents prefer quality information, but have behavioral limitations in managing a heavy flow of information. We measure the relationship between the quality of an idea and its likelihood to become prevalent at the system level. We find that both information overload and limited attention contribute to a degradation in the market's discriminative power. A good tradeoff between discriminative power and diversity of information is possible according to the model. However, calibration with empirical data characterizing information load and finite attention in real social media reveals a weak correlation between quality and popularity of information. In these realistic conditions, the model predicts that high-quality information
has little advantage over low-quality information.
\end{abstract}

\end{frontmatter}

Four centuries ago, the English poet John Milton argued that in a free and open encounter of ideas, truth prevails~\cite{Milton1644Areopagitica}. Since then the concept of a \emph{free marketplace of ideas} has been used to support free speech policies and even applied to the study of scientific research~\cite{ZamoraBonilla2012823}. The theory draws analogies to natural selection, where the traits of a species determine its survival, and to economic markets, where the intrinsic value of a good determines its success. Two necessary elements of this theory are the \emph{diversity} of ideas to which people are exposed and the \emph{discriminative power} of the marketplace, which we define as its ability to allow better ideas to become more popular. 

The recent advent of social media as a major communication platform is having a significant impact on the marketplace by broadening participation and facilitating the contribution and exchange of information and opinions. We use the terms ``idea'' and ``meme'' interchangeably to mean a transmissible piece of information~\cite{dawkins1989selfish}. A meme can represent a link to a news article, a phrase, a hashtag, or a video or image. Through networks such as Twitter and Facebook, users are exposed daily to a large number of memes that compete to attain success. However, cognitive constraints limit the number of social interactions we can sustain~\cite{10.1371/journal.pone.0022656} and the number of ideas we can consider, giving rise to an ``attention economy''~\cite{simon_designing_1971, refgoldhaber, falkinger, Ciampaglia2015production}. The information flows that result from such complex dynamics have increasingly consequential implications for politics and policy~\cite{Conover2010predicting, conover12partisan, 10.1371/journal.pone.0079449}, making the questions of discrimination and diversity more important in today's online information markets than ever before. 

In this paper we study discriminative power and diversity in a model information network, similar to modern social media, where memes are shared and spread for person to person. We assume the existence of an intrinsic measure of quality for each meme shared online, and explore how two critical factors --- the number of competing memes and the finite attention of the participants --- affect the system's ability to select the best memes for survival and diffusion, while sustaining a diverse ecosystem of ideas. We observe a tradeoff between quality discrimination and diversity, in which both can be relatively high when agents have sufficient attention and are not overloaded with information. Calibrating the model to empirical data is difficult because qualities such as value, innovation, reliability, and relevance of information can rarely be measured or even defined \textit{a priori,} making it difficult to quantify the discriminative power of online information markets. However, it is possible to characterize the distributions of information load and attention from empirical social media data. Unfortunately, these measurements place online information networks in a regime where quality and popularity of information are weakly correlated, far from the optimal tradeoff. 

Our work draws inspiration from previous work, both in economics and social science, that has studied the relation between ``quality" and popularity from both theoretical and empirical perspectives. Adler~\cite{Adler85} has shown that simple rationality arguments based on the cost of learning about quality will lead to ``stars'' with disproportionate popularity even in the absence of differences in quality. Here we assume that the cost of learning is zero; every agent can evaluate quality, although there is noise in the choice. Weng \textit{et. al.}~\cite{refweng} demonstrated that some memes inevitably achieve viral popularity irrespective of quality in the presence of competition among networked agents with limited attention. Their model did not incorporate an individual preference for quality memes. In their seminal ``music lab" experiment~\cite{Salganik854}, Salganik \emph{et al.} have shown how the relation between quality and popularity can be distorted by introducing mechanisms that allow ``consumers" to choose knowing about the aggregated choices of their peers. We instead assume a networked system where new items to choose from are continuously introduced and where agents only have access to local information shared by their neighbors. We focus on the capacity of the system to let quality emerge, how this discrimination is affected by the cognitive limitations of individual agents, and how it affects the level of diversity the system can support. 

The simple model presented here does not \emph{explicitly} incorporate many behavioral, social, and technological mechanisms that affect discrimination and diversity in the online marketplace of ideas. For example, prior research has studied the role of technology in discriminating between truthful information and misinformation. It has been argued that truth is easier to distribute because one can more easily verify truthful sources, making online misinformation expensive to sustain~\cite{SnowBailard}. However, the ease of disseminating misinformation through social media may counter this argument. The ``wisdom of the crowd'' enabled by social media~\cite{surowiecki2005wisdom} should also facilitate the discrimination of information based on quality by combining the diverse opinions of many individuals~\cite{page2008difference}. But when people communicate, their opinions are no longer independent, leading to higher confidence and lower accuracy~\cite{Lorenz2011}. 

Cognitive and behavioral processes for dealing with opinions that challenge one's beliefs may decrease our capability to discriminate between high- and low-quality information~\cite{Collins2011,Smith2009Contextualizing}. For example, conformation bias~\cite{nickerson1998confirmation} may have evolved as an effective strategy to avoid misinformation, by comparing incoming information to one's own existing beliefs, and adopting it if it is sufficiently concordant~\cite{Smith2014Evil}. However, in social media, such a bias easily leads to ineffective discrimination; strategies such as accepting new information if it comes from multiple sources~\cite{Centola2007} are not useful because people lack knowledge of the social network structure necessary to determine whether multiple information sources are independent of each other.  
Confirmation bias may be reinforced online by our limited capacity to cope with the information overload caused by the messages that flood our screens \cite{hiltz1985structuring} and our consequent need to quickly discard irrelevant information.  

The diversity of information present in the market can also be affected by the interplay between behavioral and cognitive factors and algorithmic biases of online social networks. It is easy to rewire our connections and affect the sources of information to which we are exposed~\cite{frey1986recent}. Mechanisms such as triadic closure, facilitated by social media recommendation, may be suboptimal for the discovery of relevant but unfamiliar information~\cite{Weng:2013:RID:2487575.2487607,Babaei:2016:EIN:2835776.2835826}. These selection processes may cluster people into a few homogeneous factions~\cite{Axelrod1997}, often called ``echo chambers'' \cite{Sunstein1} or ``filter bubbles'' \cite{Pariser}. This may further lead to polarization \cite{Sunstein2,Truthy_icwsm2011politics,stanovich2013,Nikolov15socialbubbles}; one group may automatically discount ideas from another \cite{Mason2007,Nisbett1991}.

This body of work suggests that, paradoxically, our behavioral mechanisms to cope with information overload may make online information markets less meritocratic and diverse, increasing the spread of misinformation \cite{Nyhan:2010aa, DelVicario19012016} and making us vulnerable to manipulation \cite{Truthy_icwsm2011class, socialbots-CACM}. Anecdotal evidence of hoaxes, conspiracy theories, and fake news in online social media is so abundant that massive digital misinformation has been ranked among the top global risks for our society \cite{Forum:2013}. And fake news have become a major topic of debate in the US and Europe.

Several studies have investigated the role played by network mechanisms affecting the popularity of individual memes. 
Crane and Sornette \cite{refcrane} proposed an epidemic model on a social network to describe the exogenous and endogenous bursts of attention toward a video. 
Ratkiewicz \textit{et al.}~\cite{ratkiewicz} employed a model in which random collective shifts of attention due to exogenous events provide a way to interpret the broad distribution of magnitude in popularity bursts.
Bingol~\cite{refbingol} proposed a dynamic model where agents can remember and forget, and use recommendation to discover new agents. This model predicts that the popularity of an agent is linearly related to memory size. 
Huberman~\cite{refhuberman} studied the effects of the content's novelty and popularity in attracting attention. Wu and Huberman~\cite{refwu} developed a  model with the novelty of a news story fading with time and showed that attention decays over a natural time scale.  Lerman and colleagues~\cite{Hodas12socialcom,Kang15sbp} showed that the combination of competition and position bias (a manifestation of limited attention in social media) affects the visibility of a meme and thus constrains social contagion.

The above literature considers the popularity of pieces of information in isolation. Markets in which \emph{many} memes compete for the limited attention of social media users have received scarce consideration. A notable exception is the work of Weng \textit{et. al.}~\cite{refweng}, who used an agent-based model to demonstrate that the combination of social network structure and finite attention of social media users are sufficient conditions for the emergence of viral memes.
Gleeson \textit{et. al.}~\cite{refgleeson,Gleeson_PRX} formalized this model as a critical branching process, predicting that the popularity of memes follows a power-law distribution with very heavy tails. 

These results reveal that quality is not a necessary ingredient to explain popularity patterns in online social networks, but say nothing about the actual importance of information quality. It is reasonable to assume that quality does play a role in individual decisions about information consumption. 
This motivates our theoretical 
analysis to determine whether discrimination of information according to its quality at the individual level can be reflected in discriminative power at the system level, and at what cost in terms of the market's capability to sustain diversity of information. 


\paragraph{{\bf Model of information sharing network}}

\begin{figure}
\centerline{\includegraphics[width=0.5\textwidth]{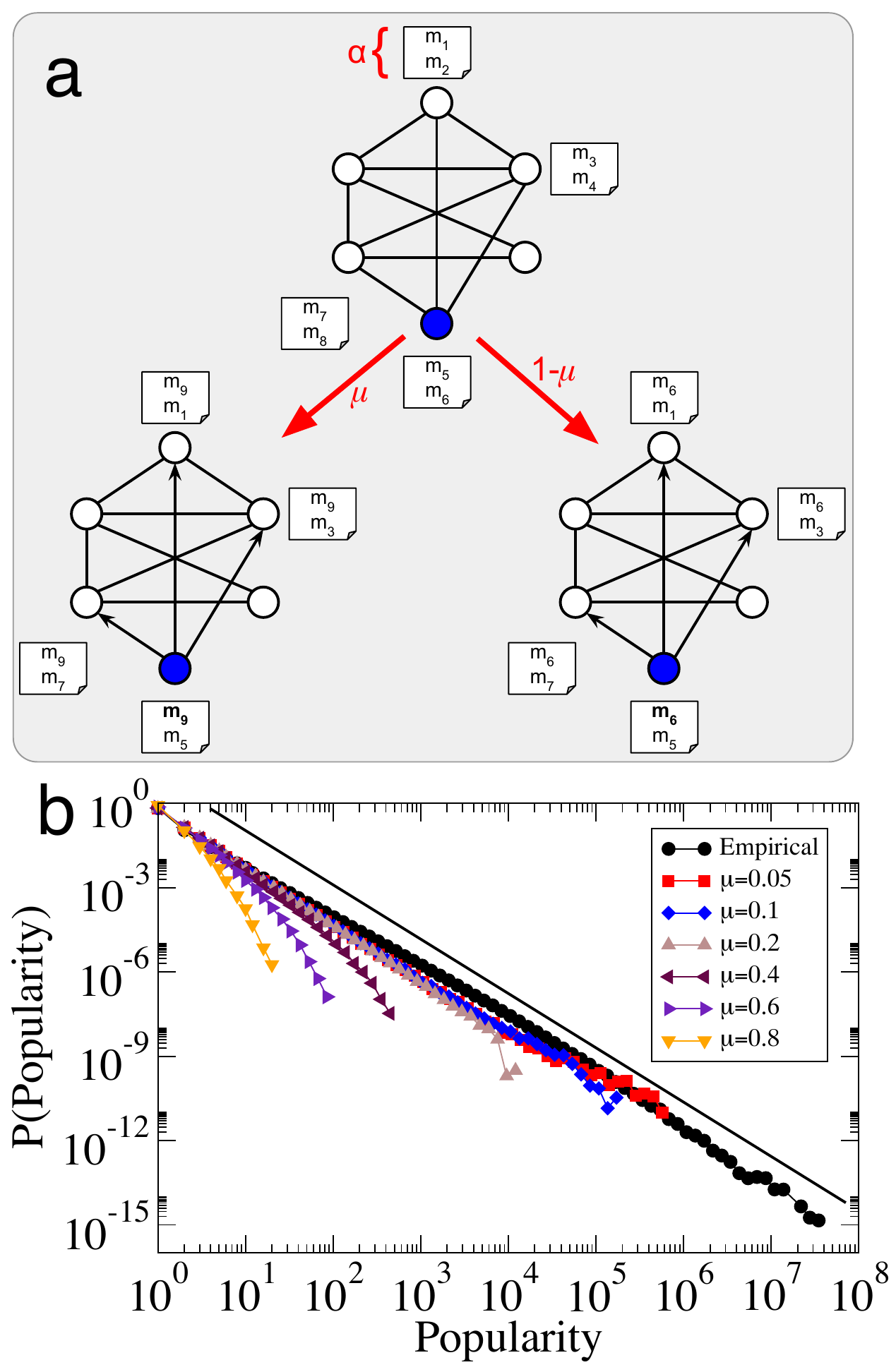}}
\caption{\textbf{Illustration of the meme diffusion model and predicted popularity distributions.} (a) At each time step, an agent is considered (shaded). The agent chooses to create and share a new meme ($m_9$) with probability $\mu$.  Otherwise, with probability $1-\mu$, the agent reshares one of $\alpha$ messages in its feed, to which it is currently paying attention ($m_6$). The message is transmitted to the agent's neighbors and appears on the top of their feeds. 
(b) Distribution (PDF) of meme popularity $p$. We compare the model predictions ($\alpha=10$) with an empirical distribution obtained by counting the number of occurrences of hashtags in a sample of public tweets. This empirical popularity is distributed according to a power-law distribution $P(p) \sim p^{-1.94}$ (see guide for the eye).
\label{fig:panel5}}
\end{figure} 

We aim to examine the conditions in which the ``best'' ideas are those that capture a greater portion of collective attention, and whether this happens at the expense of the diversity of ideas.  To this end, we propose a simple agent-based model inspired by the long tradition of representing the spread of ideas as an epidemic process where messages are passed along the edges of a network \cite{morris,goffman,daley,bailey,goetz}. Agents are represented by the nodes of a static network where the links embody social connections. The network dynamics in the model capture the salient ingredients common to  popular social media platforms. Each message, or post, carries a ``meme'' or ``idea,'' i.e., the unit of information that spreads from person to person~\cite{dawkins1989selfish}. Different messages may carry the same meme. {Fig.~\ref{fig:panel5}(a) illustrates the dynamics of the model.} 

To examine the discriminative power of the market, we imagine that each meme is characterized by an intrinsic \emph{quality} value.  
Agents pay attention to memes shared by their neighbors. We assume that the probability that an agent shares one of these memes, allowing it to spread, is proportional to the meme's quality. 
The quality might represent different properties that make the meme more likely to be shared, depending on the situation being modeled: the originality of an idea, the beauty of a picture, and the truthfulness of a claim are valid examples. 

In contrast with classical epidemiological models, messages carrying new memes are continuously introduced into the system in an exogenous fashion. We use the rate $\mu$ at which this happens as a parameter of the model to regulate the \emph{information load} of the agents, i.e., the average number of memes received by an agent per unit time.

Agents produce messages containing new memes and reshare messages originated or forwarded by their neighbors. When resharing, an agent is capable of paying \emph{attention} to only a finite number $\alpha$ of messages at a time. If we think of messages from neighbors as appearing in, say, reverse chronological order on a social media feed, a user during a session will scroll down the feed to view $\alpha$ recent posts. 
Further details about the model are presented in Methods.

\paragraph{{\bf Effects of information load and finite attention}}

Let us investigate how the information load affects the relationship between meme quality and success. 
The relationship is not trivial because memes are not shared on the sole basis of their quality; a meme with lower quality may be selected if it is over-represented in an agent's feed. The actual probability that a meme is shared depends on a complex interplay of factors that include its current popularity, the network structure, and the limited attention of the agents. 
There are multiple ways to define the success of a meme, for instance by its longevity. Here  we measure its popularity, defined as the number of times the meme is shared across the network from the moment it is first injected until it finally disappears. 
{The distribution of meme popularity, shown in Fig.~\ref{fig:panel5}(b), depends on $\mu$. For high $\mu$, the distribution is exponentially narrow and no memes go viral. As the information load becomes lighter ($\mu < 0.2$), our model reproduces the broad distribution from the empirical data, indicating that a few memes spread virally through the population. In the absence of quality, we would expect a power-law distribution of popularity $P(p) \sim p^{-\beta}$ with exponent $\beta \approx 1.5$~\cite{refgleeson,Gleeson_PRX}. However, fitting~\cite{clauset2009power} reveals a larger exponent $\beta \approx 1.94$. This is consistent with a model of a branching process with uniform fitness, which predicts an exponent $\beta = 2$~\cite{ASI:ASI20653}.}

\begin{figure}[t]
\centerline{
 \includegraphics[width=\textwidth]{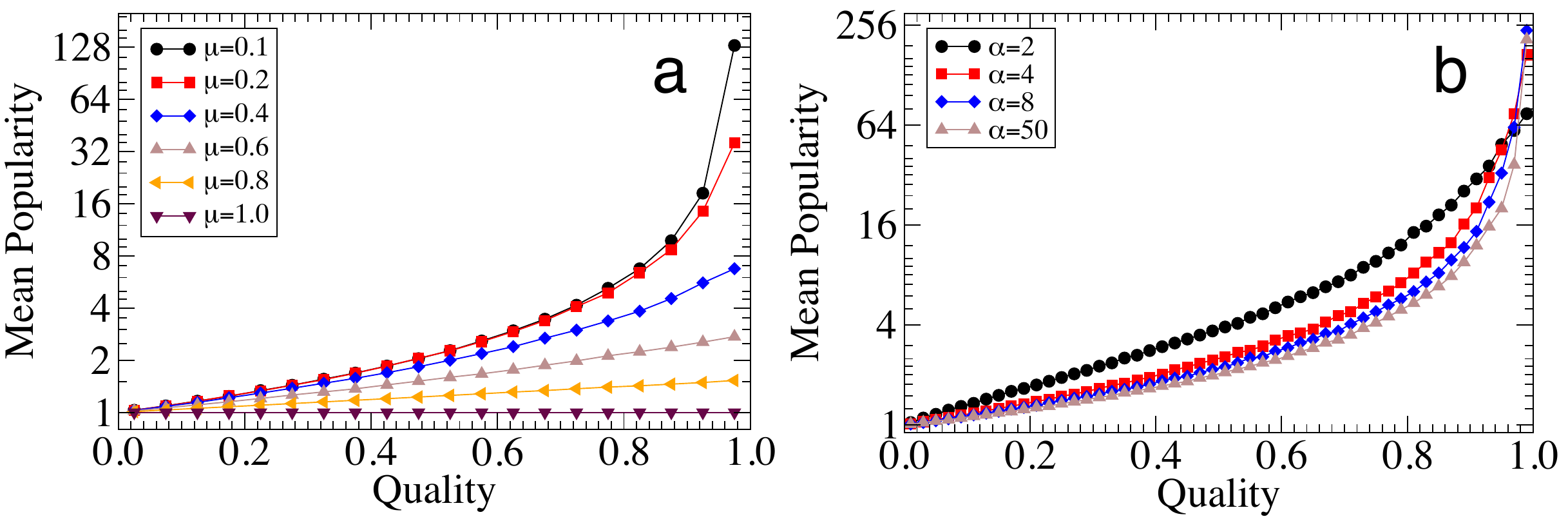}
}
\caption{\textbf{Average popularity of memes.} The number of times a meme is shared is plotted as a function of quality for different values of (a) information load $\mu$ ($\alpha=10$) and (b) attention $\alpha$ ($\mu=0.1$).
\label{fig:popVfit}}
\end{figure} 

{The relationship between meme quality and popularity is illustrated in Fig.~\ref{fig:popVfit}(a). On average, memes with higher quality do have a better chance to survive and succeed, but in a way that depends considerably on $\mu$.} A single meme survives in the limit $\mu=0$, typically but not necessarily one with high quality. For small values of $\mu > 0$, very high quality yields a disproportionally large chance to succeed. For large $\mu$ ($\mu < 1$), the relative advantage conferred by a higher quality is much smaller. In the limit case $\mu=1$, a new meme is introduced at all times and therefore there is no chance for memes to spread, irrespective of quality. In summary, an increase in information load corresponds to a decrease in discriminative power because quality has a lesser effect on popularity. 

The effect of finite attention on the relationship between quality and popularity is illustrated in Fig.~\ref{fig:popVfit}(b). 
We assume $\alpha > 1$ so that agents have some choice in selecting which memes to share. As expected, the mean popularity grows with quality: the best memes have much higher chances to win. As the amount of individual attention $\alpha$ increases, the curves become more concave; the mean popularity grows more slowly except for the highest values of quality, an indication of increased selective pressure favoring the best memes. 

\begin{figure*}
\centerline{\includegraphics[width=0.85\textwidth]{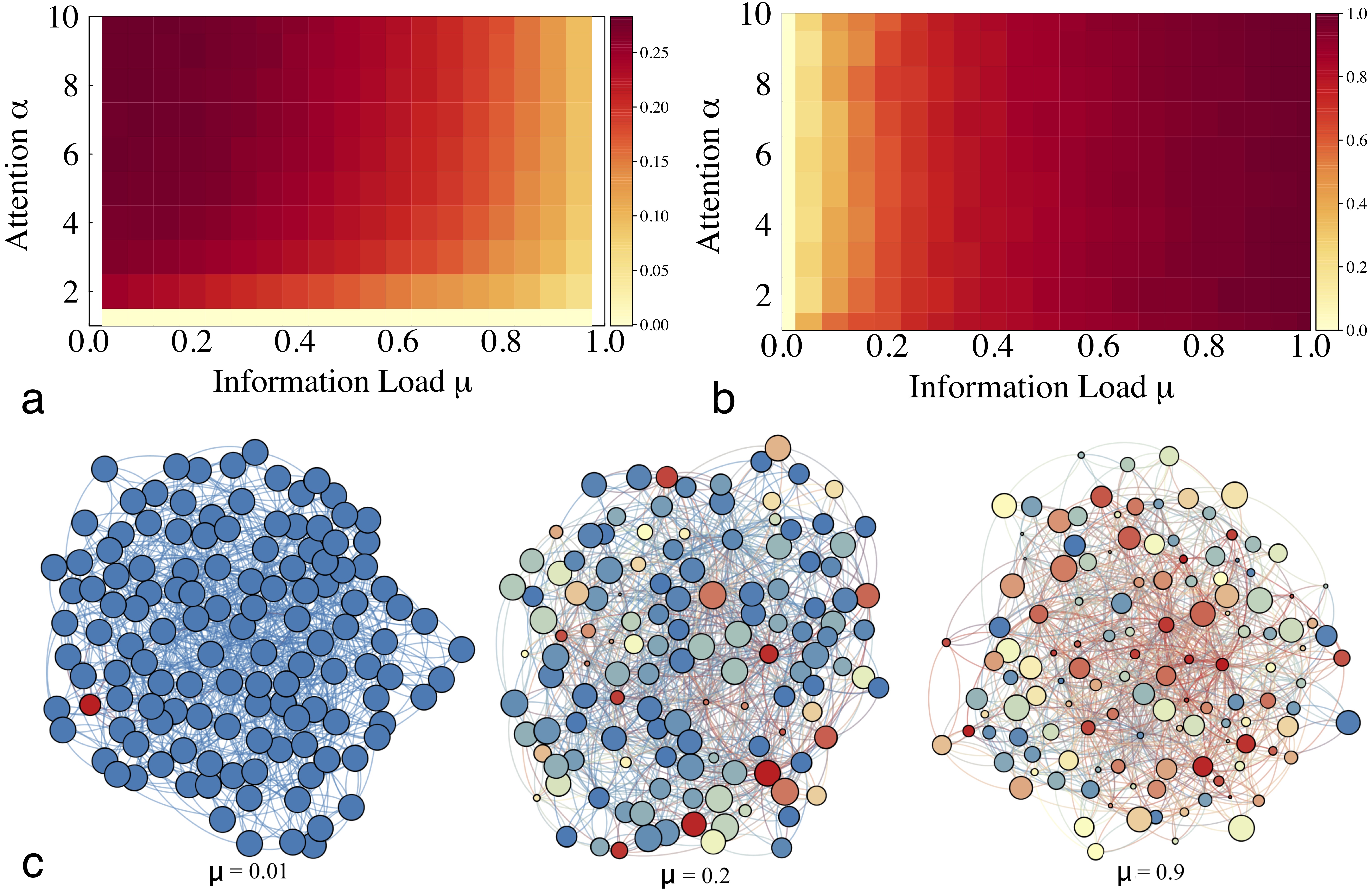}} 
\caption{\textbf{Discriminative power and diversity.} (a) Discriminative power $\tau$ as a function of information load and finite attention. (b) Diversity $H/H(\mu=1)$ as a function of intensity of information load and attention. (c) Illustration of the tradeoff between discriminative power of the system in spreading quality memes and diversity of content in the network. Nodes represent agents, their color represents the last shared meme, and their size indicates the quality of that meme. When the information load $\mu$ is small, only high-quality memes are present, with low diversity. As $\mu$ increases we observe higher diversity and lower discriminative power. Here $N=128$ and $\alpha=10$.
\label{fig:panel2}}
\end{figure*} 

We can summarize the dependency between the quality of memes and their success in a single \emph{discriminative power} measure by looking at the correlation between quality and popularity. Since the two quantities are not normally distributed, we employ the Kendall rank correlation coefficient $\tau$, which is computed by ranking memes according to the two criteria and then counting the number of meme pairs for which the two rankings are concordant or discordant, properly accounting for ties~\cite{Kendall1938}. High $\tau$ indicates that fitter memes are more likely to win, granting the system discriminative power; in the extreme case $\tau=1$ the two rankings are completely concordant. Small $\tau$ signifies a lack of quality discrimination by the network. Fig.~\ref{fig:panel2}(a) shows that network discriminative power degrades both with higher information load and with more limited attention. 
Similar results are obtained when using mutual information in place of Kendall's $\tau$ to measure discriminative power. 

\paragraph{\bf Tradeoff between diversity and discriminative power}

As discussed above, discriminative power in spreading quality content is a desirable property of a social network. A second desirable property of an ideal communication system is the preservation of information diversity, i.e., the possibility to have many distinct memes alive simultaneously. As illustrated in Fig.~\ref{fig:panel2}(c), the two goals are in contradiction --- the price associated with the capability of the network to let a high-quality meme prevail is a loss in diversity, with many memes receiving relatively small attention despite their intrinsic quality.
Let us therefore explore the tradeoff that results from the competition for attention in the network. 

To measure the amount of diversity in the system at the steady state, we start from the entropy $H = -\sum_m P(m) \log P(m)$ where $P(m)$ is the portion of attention received by meme $m$, i.e., the fraction of messages with $m$ across all of the user feeds. The sum runs over all memes present at a given time and is averaged over a long period after stationarity has been achieved (see Methods). The minimum entropy is zero, when all nodes have the same meme $(\mu=0)$. The maximum entropy, 
obtained in the extreme case $\mu=1$, 
depends on $\alpha$.  
To discount this dependence we measure diversity by the normalized entropy $H/H(\mu=1)$. Fig.~\ref{fig:panel2}(b) shows that with this normalization, the diversity does not depend in a significant way on the attention $\alpha$.
As expected, the diversity increases with information load and is maximized for high $\mu$.

\begin{figure}[t]
\centerline{\huge{a}\includegraphics[width=0.5\textwidth]{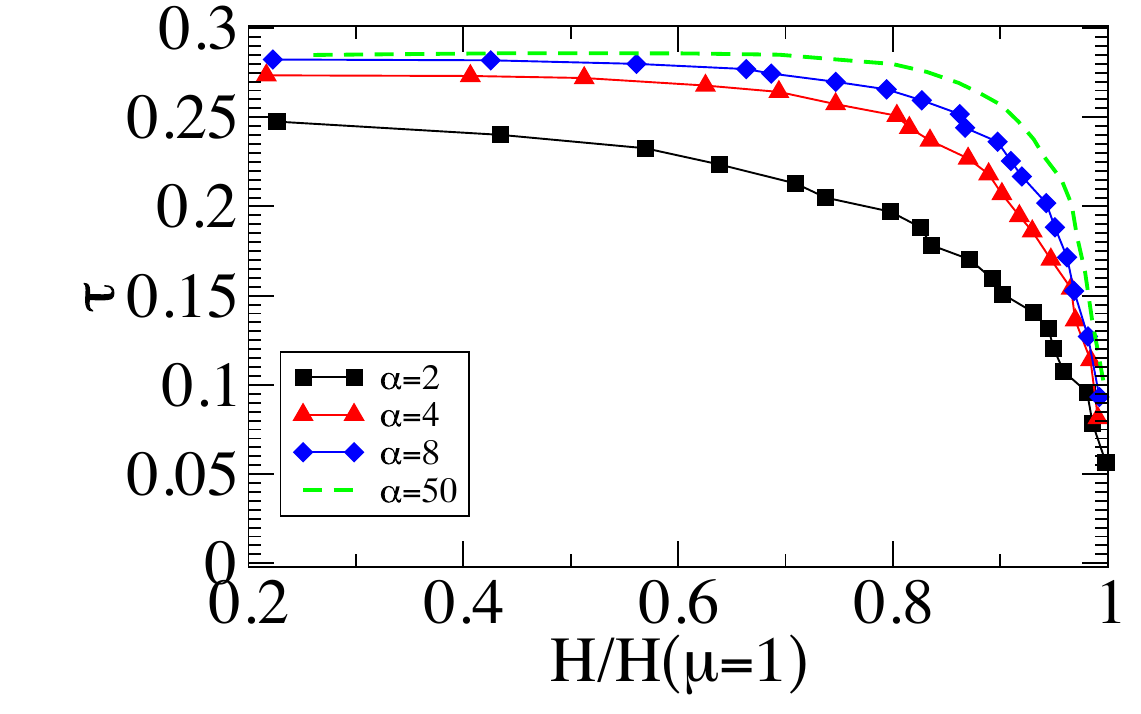}
b\includegraphics[width=0.5\textwidth]{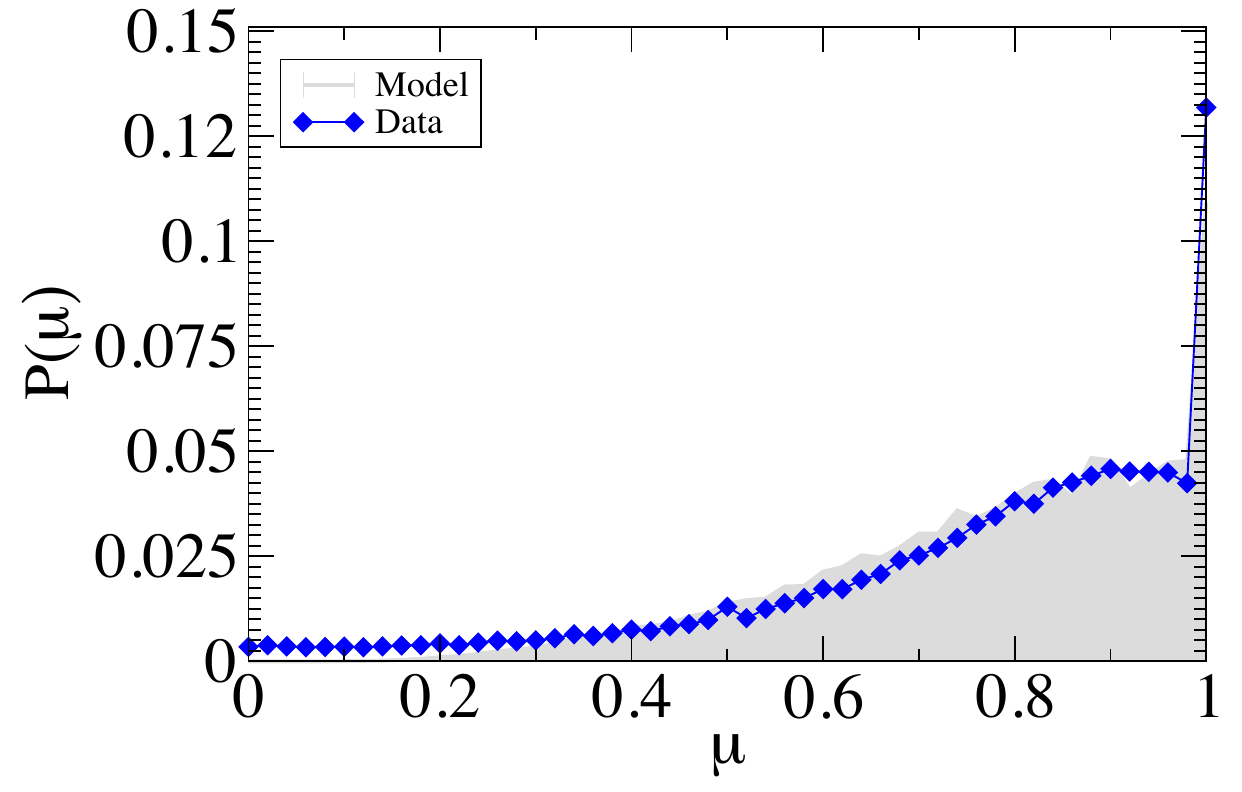}}
\centerline{\huge{c}\includegraphics[width=0.5\textwidth]{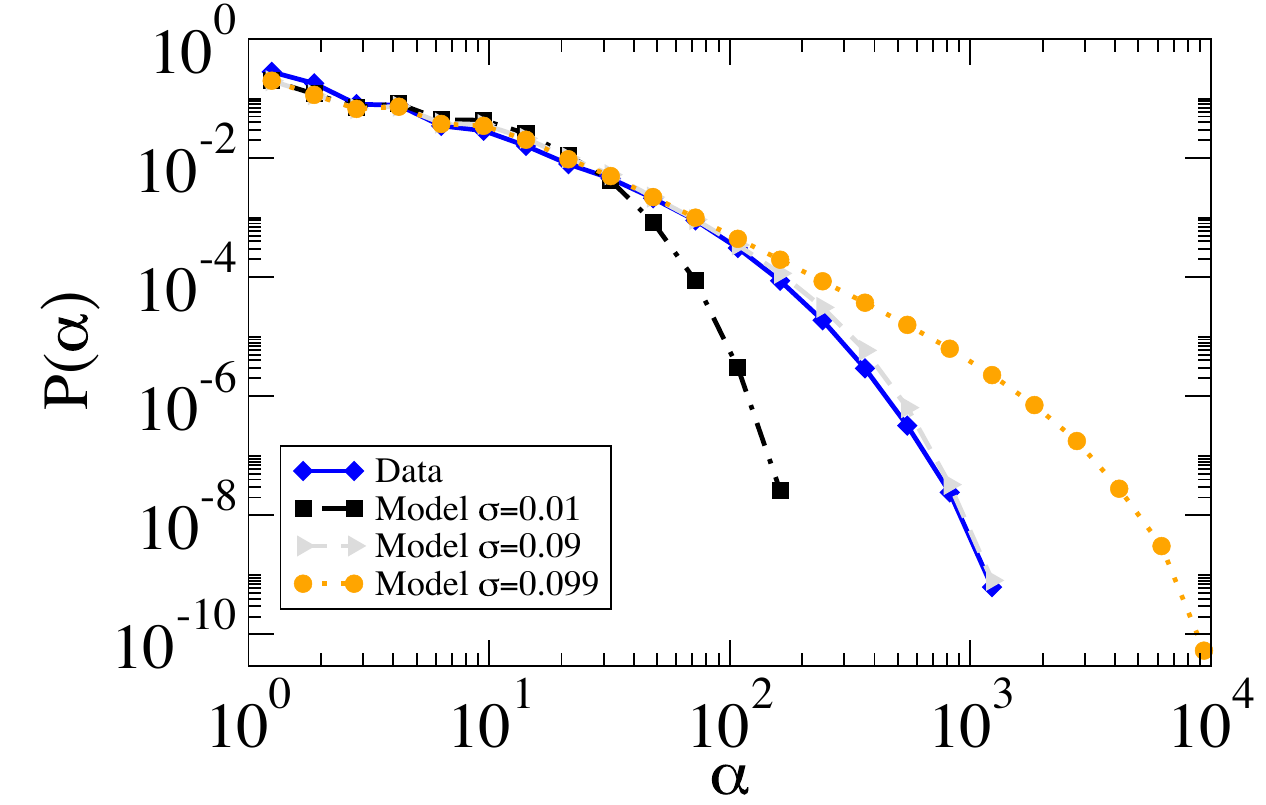}
d\includegraphics[width=0.5\textwidth]{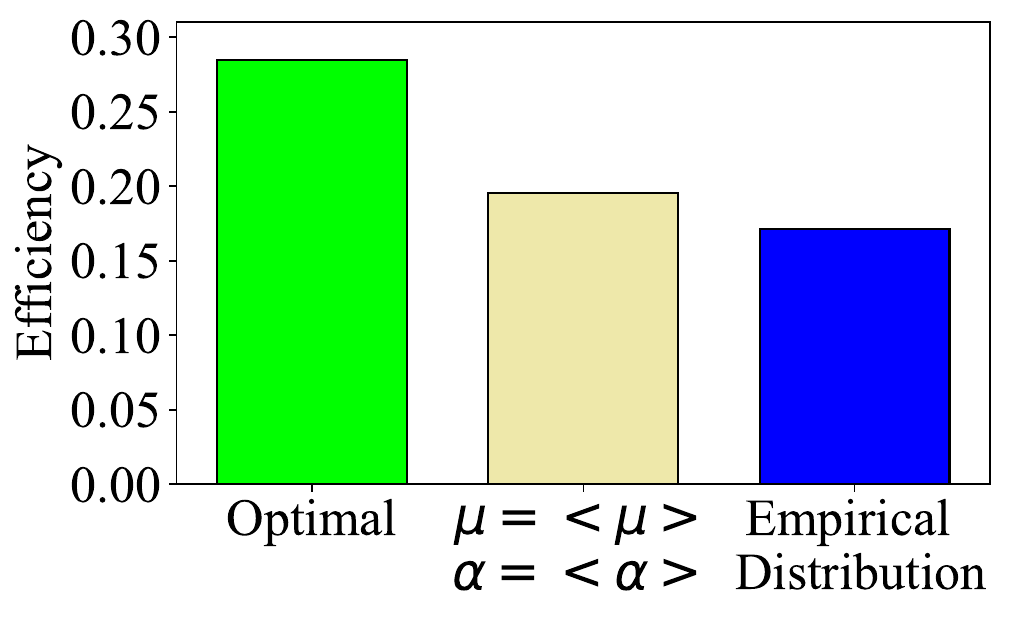}}
\caption{\textbf{Discriminative power: tradeoff with diversity and empirical calibration.} (a) Tradeoff between discriminative power $\tau$ and diversity $H/H(\mu=1)$. For each value of the attention $\alpha$, 
different tradeoffs are obtained by varying $\mu$ between 0 and 1. 
The dashed line suggests convergence in the limit for high $\alpha$. 
(b) Empirical distribution of $\mu$ measured from Twitter. The solid line is produced by the scrolling model described in Methods, with $\rho = 0.978$, $\langle q \rangle = 0.1$, and $\sigma = 0.09$.
(c) Empirical distribution of attention $\alpha$ derived from Tumblr.  
The lines are produced by the scrolling model described in Methods, with $\rho = 0.05$ and $\langle q \rangle = 0.1$.
(d) Discriminative power $\tau$ of the system for different empirical calibrations of information load $\mu$ and attention $\alpha$. The optimal case is obtained for small information load ($\mu \leq 0.5$) and large attention ($\alpha \geq 50$). 
\label{fig:panel3}}
\end{figure} 

The tradeoffs between discriminative power and diversity are better illustrated in Fig.~\ref{fig:panel3}(a). For any value of finite attention $\alpha > 1$ we observe a transition from relatively high discriminative power and low diversity (when information load is low) to high diversity and low discriminative power (high information load). The amount of attention $\alpha$ has a significantly effect on the tradeoff: for a given level of diversity the discriminative power improves when people can pay attention to multiple memes, and vice versa the network can sustain a larger diversity without loss in discriminative power. When $\alpha$ is large, there is a region where the network can sustain very high diversity with relatively small loss in discriminative power. 

\paragraph{\bf Empirical calibration}

The model has two key parameters, the information load $\mu$ and the individual attention $\alpha$. We have made two simplifying assumption about these parameters: that agents in our model introduce new memes at the same rate $\mu$ and have the same amount of attention $\alpha$. In the real world, some people may post more new memes, others may tend to reshare memes adopted through their connections; some may pay attention to only a handful of messages, others may scroll through social media feeds for prolonged periods. Let us turn to empirical data to calibrate these ingredients of the model. 

We counted posts and reposts by a large sample of social media users to estimate the portions of original and reshared memes per user (see Methods). Fig.~\ref{fig:panel3}(b) shows the resulting empirical distribution of the rate of introduction of new memes, which corresponds to the parameter $\mu$ in our model. The information load spans the entire spectrum ($\mu \in [0,1]$) but is skewed toward high values with a peak at $\mu = 1$, corresponding to users who post but do not reshare.  The average is quite high ($\langle \mu \rangle \approx 0.75$). 

The amount of attention one devotes to assessing information, ideas and opinions encountered in online social media varies not only across persons but also depending on time and circumstance; the same user may be hurried one time and careful another. We counted the number of times that a user stops on a post during a scrolling session on a social blogging platform to estimate their finite attention (see Methods). This number corresponds to the parameter $\alpha$ in our model. Fig.~\ref{fig:panel3}(c) shows that the resulting empirical distribution of $\alpha$ is broad, with average 
$\langle \alpha \rangle \approx 14$. 

A naive way to take these data into account for calibrating our model market of ideas is to set the information load and attention parameters to their empirical averages, $\langle \mu \rangle \approx 0.75$ and $\langle \alpha \rangle \approx 14$, respectively. This yields the value of discriminative power $\tau$ shown in Fig.~\ref{fig:panel3}(d), which is roughly 70\% of the maximum $\tau$ obtained in the limit of high $\alpha$ and low $\mu$. 
We can easily remove the simplifying assumption about constant $\mu$ and $\alpha$ using insight from the data.  
We therefore adopted a second calibration of the model by drawing 
both $\mu$ and $\alpha$ from the empirical distributions of information load and attention. We repeated the analysis with this more realistic model and found a significantly lower value of discriminative power, $\tau \approx 0.17$ (Fig.~\ref{fig:panel3}(d)). 
This finding suggests that the heterogeneous information load and attention of the real world lead to a market that is incapable of discriminating information on the basis of quality. 

Given the importance of the heterogeneity of information load and attention across users, we investigated the possibility of reproducing their empirical distributions as an outcome of information market mechanisms. We extended the model by taking into account how users scroll through their social media feeds. The scrolling model, illustrated in Methods, reproduces the empirical distributions of $\mu$ and $\alpha$ as well as the decrease in discriminative power.

\paragraph{\bf Virality of low-quality information}

\begin{figure}[t]
\centerline{\includegraphics[width=0.8\textwidth]{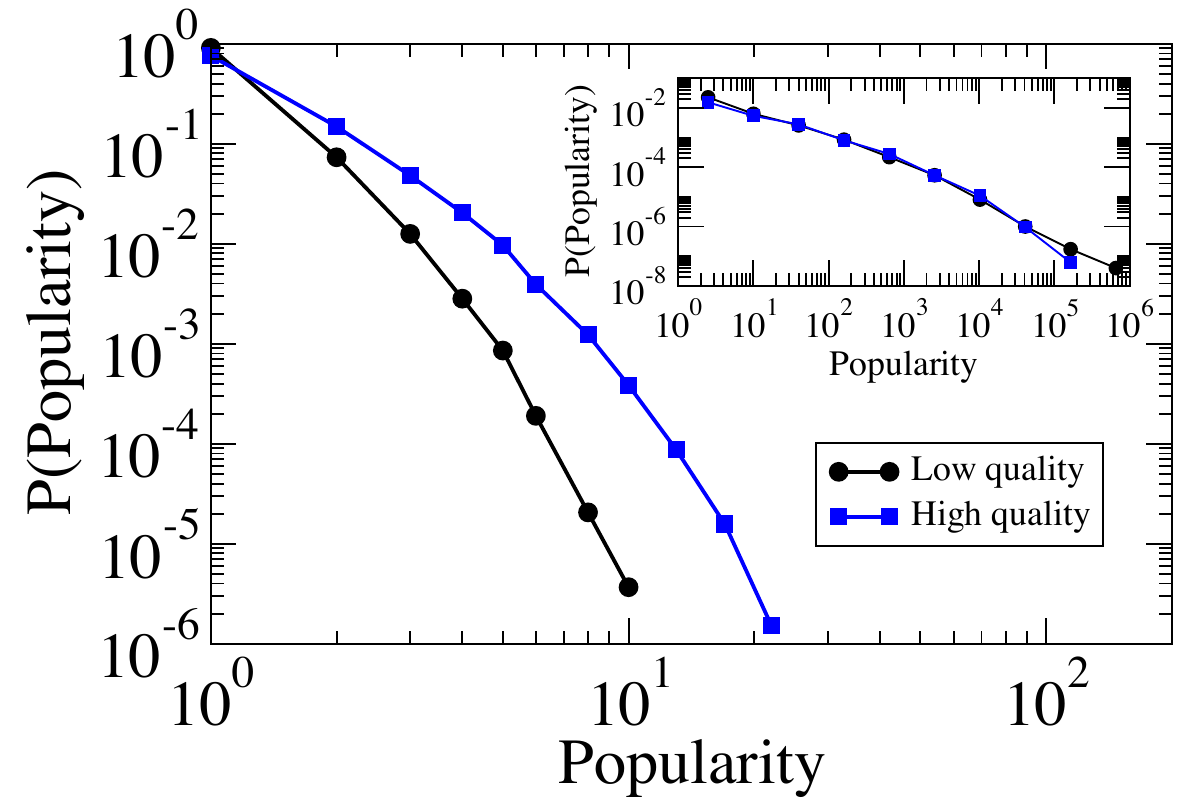}}
\caption{\textbf{Popularity distributions.} Distribution of the number of times that memes in two categories are shared in our model. The top two thirds of memes are considered high-quality ($f \geq 0.34$) and the rest are considered low-quality. Inset: Distribution of the number of times that news articles in two categories are shared on Facebook: articles that support false claims or undermine true claims (low quality) and those that fact-check false claim or support true claims (high quality). 
\label{fig:pop_distr_by_fitness}}
\end{figure}

The low $\tau$ predicted by our model means that the system is unable to discriminate between low-quality and high-quality information. Fig.~\ref{fig:pop_distr_by_fitness} illustrates this finding by plotting the distributions of popularity for two groups of low-quality and high-quality memes, respectively. We observe that high-quality memes have little competitive advantage in terms of their chances of success.

Empirical validation of this model prediction requires the identification of some feature that translates into a quality metric for information that is shared online. While this is generally difficult, proxy measures of quality exist in some cases. We used empirical data from Emergent (see Methods) about posts shared on social media with links to news articles in two groups. Articles in one group support claims debunked by fact-checkers or undermine claims verified by fact-checkers. The other group includes articles that fact-check hoaxes or support verified claims. It is reasonable to assume that most people would consider articles in the second group as having higher quality than those in the first. 
We compared the numbers of times articles in these two groups were shared online. As illustrated in the inset of Fig.~\ref{fig:pop_distr_by_fitness},  the articles in the two groups are just as likely to go viral.

\paragraph{\bf Discussion}

The proposed model is quite minimal and relies on few parameters, but it captures salient behavioral features that shape the diffusion of information in online social networks. This allows us to study how information load and limited attention affect the discriminative power of the network, i.e., the likelihood that the best memes will succeed at reaching many people. Our main finding is that the survival of the fittest is far from a foregone conclusion where information is concerned. 
Both information load and limited attention lead to low discriminative power, so that it becomes very difficult for the best memes to win. 
Meme diversity can coexist with network discriminative power when we have plenty of attention and are not overloaded with information.

One important question that deserves further exploration is the role played by the network structure in determining market discriminative power and diversity. While the results presented here are robust to changes in network size, density, and clustering (see Methods), the model could be further expanded to capture other characteristics derived from empirical social networks, such as the segregated communities that we typically observe around discussions of polarizing topics~\cite{Truthy_icwsm2011politics, conover12partisan}. How the predictions of our model depend on these features remains to be investigated.

Empirical validation of the predictions generated by our model remains a challenge, given the difficulty to quantify the factors that affect the intrinsic quality of a meme in the real world. However, we have shown that it is possible to derive empirical estimations of the key parameters in our model, namely the rate of introduction of new memes and the depth of user attention. According to these calibrations, real social media have heterogeneous levels of information load and attention, which place them in a regime of low discriminative power. If prior research had revealed that intrinsic quality is not a \emph{necessary} ingredient to explain the broad distribution of meme popularity in social media~\cite{refweng,refgleeson}, the present results are not much more reassuring. They suggest that better memes do not have a significantly higher likelihood to become popular compared to low-quality information. The observation that hoaxes and fake news spread as virally as reliable information in online social media (Fig.~\ref{fig:pop_distr_by_fitness} inset) is not too surprising in light of these findings.

Our results suggest that one way to increase the discriminative power of online social media would be to reduce information load by limiting the number of posts in the system. Currently, bot accounts controlled by software make up a significant portion of online profiles~\cite{varol:2017:icwsm}, and many of them flood social media with high volumes of low-quality information to manipulate public discourse~\cite{socialbots-CACM,FM7090}. By aggressively curbing this kind of abuse, social media platforms could improve the overall quality of information to which we are exposed.

\subsection*{Methods} 

\paragraph{{\bf Diffusion model and simulation details}}

{The basic setting for our model is a set of agents connected by a social network. Each agent holds a feed of the $\alpha$ most recent messages produced by their neighbors.} The reverse chronological ordering of the feed is a realistic simplifying assumption, which is accurate in social media platforms such as Twitter. In some cases, such as Facebook, the ranking algorithms also considers factors like popularity and social engagement. However, all platforms give strong priority to recent messages.

At each time step one agent $i$ is chosen at random. With probability $\mu$, $i$ produces a message carrying a new meme. The meme's \emph{quality} is drawn uniformly at random from the unit interval. Alternatively, with probability $1-\mu$, $i$ selects one of the messages in its feed. The probability that an agent selects a specific message from its feed is proportional to the meme's quality. More explicitly, let $M_i$ be the feed of $i$ ($|M_i|=\alpha$). The probability of message $m \in M_i$ being selected is $P(m) = f(m) / \sum_{j \in M_i} f(j)$ where $f(m)$ is the quality of the meme carried by $m$. The message is added to the feeds of $i$'s neighbors; if a feed exceeds $\alpha$ messages, the oldest is forgotten. 
This mechanism represents how finite attention is allocated to information posted by one's social connections. 

The two parameters of the model allow us to explore how the intensity of the information load ($\mu$) and the attention depth ($\alpha$) interact with the intrinsic value of an idea and affect its chances to win. 

We analyze the behavior of the model by simulating the diffusion and information load process on synthetic scale-free networks (see below). 
As the competition takes place, some of the memes die fast while others live longer and infect a large fraction of the network. Such a process continues until the systems reaches a \emph{steady state} in which the average number of distinct memes remains roughly constant. This number depends on $\mu$. 

{The popularity of a meme can be defined by the cumulative attention it gathers across the network. In practice, we measure popularity by counting the number of times a meme is shared or reshared. The measurements occur at steady state.} 

For each experiment and set of parameter values, we simulated the dynamics of the model on 
a synthetic undirected network. While some popular social networks are directed, many connections are reciprocal~\cite{Kumar:2006,Kwak:2010}. 
The results presented here employ scale-free networks built with the preferential attachment model, with $N=10^3$ nodes and average degree $\langle k \rangle = 20$. The results generalize to larger and sparser networks with higher clustering coefficient~\cite{PhysRevE.65.026107}. 

Once the system reached the steady state, we performed measurements to determine the success of a meme. To this end we considered only memes that were introduced after the system reached the steady state. We followed each of these memes from the moment it was first shared until it completely disappeared from the network, recording its quality as well as its popularity. 
During each simulation, we monitored 100,000 memes that were introduced and forgotten after the system reached the steady state. We ran each simulation 20 times, so that our analyses of popularity took $2 \times 10^6$ memes into consideration. Measurements of discriminative power and diversity were averaged across runs. 

\paragraph{\bf Scrolling model of attention}

Rather than assuming a fixed feed depth and a single post or repost per session, imagine that users scroll through their feeds by paying attention to messages and resharing them in sequence, until they decide to stop the session. With some probability $\rho$ the user posts a message with a new meme and stops. Otherwise, with probability $1-\rho$, the user performs a scrolling session. After resharing a message, the user stops with probability $q$. Otherwise, with probability $1 - q$, the user scrolls down to view and reshare another message, and so on. Let us further assume that for each session, $q$ is drawn uniformly from an interval $[\langle q \rangle - \sigma, \langle q \rangle + \sigma]$. The parameter $\sigma$ represents the level of attention heterogeneity across scrolling sessions. 
It can be shown analytically that when $\sigma$ is small, the distribution of $\alpha$ is well approximated by an exponential decay;
in the limit $\sigma \rightarrow 0$, $P(\alpha) \sim e^{\lambda (\alpha-1)}$ with $\lambda = \ln(1 - \langle q \rangle) < 0$. 
As $\sigma \rightarrow \langle q \rangle$, for large $\alpha$ the distribution approaches a power law $P(\alpha) \sim \alpha^{-2}$; the heavy tail indicates that users occasionally scroll through a large number of messages. These behaviors are illustrated in Fig.~\ref{fig:panel3}(c).
The parameters $\rho$, $\langle q \rangle$, and $\sigma$ can be tuned to fit the empirical distributions of $\mu$ and $\alpha$ from Twitter and Tumblr, respectively (Fig.~\ref{fig:panel3}(b,c)). Both distributions can be fit by approximately the same values of $\langle q \rangle$ and $\sigma$, while $\rho$ is platform-dependent. 

To explore the effect of the attention heterogeneity $\sigma$ on discriminative power, we incorporated the scrolling mechanism into the model to generate $\alpha$ and found that high $\sigma$ leads to a significant decrease in discriminative power. This is due to the fact that although $\alpha$ increases with $\sigma$ on average, large values yield little benefit in discriminative power whereas small values cause serious discriminative power losses.

\paragraph{\bf Data}

Twitter data to measure hashtag popularity was obtained from a sample of approximately 10\% of public tweets provided by the Twitter streaming API and collected in 2014. Rare hashtags have a lower chance to be represented in this sample. We extracted the empirical rate $\mu$ from $10^6$ Twitter users based on a random sample of these messages. We counted their tweets ($n_t$) and retweets ($n_r$), then measured each user's rate as $\mu = n_t / (n_t + n_r)$. While inactive users have a lower chance to be represented in the sample, such a bias should not affect the ratio $\mu$. 

We extracted the empirical attention data from approximately $10^7$ mobile scrolling sessions observed on Tumblr during two weeks in 2016. The feed interface of this app is similar to those of other social media platforms. We consider a session ended when there is no interaction for 30 minutes or longer.
During a session, we record the number of times that a user scrolls at least 500 pixels through the feed and then stops for at least one second. This number is used as a proxy for $\alpha$. 

Data about Facebook shares of articles supporting or debunking true and false claims was collected from Emergent (\url{emergent.info}), a rumor tracking project that is no longer active. Emergent reporters manually selected and evaluated claims appearing in articles shared on social media between September 2014 and March 2015. The Emergent API provided data about 742 high-quality articles (464  supporting 56 true claims, 278 fact-checking 72 false claims) and 383 low-quality articles (10 undermining 3 true claims, 373 spreading 90 false claims). The same 2:1 ratio of high- and low-quality memes is used to select a quality threshold in the model and generate the distributions in Fig.~\ref{fig:pop_distr_by_fitness}.

\paragraph{\bf Data and code availability.}

The data and code presented in this manuscript is available at \url{github.com/fregolente/virality_of_low_quality_information}.

\paragraph{\bf Competing interests}

The authors declare no competing interests.


\paragraph{Contact author.} Correspondence and requests for materials should be addressed to Diego F. M. Oliveira. Email: \texttt{diegofregolente@gmail.com}

\subsection*{Acknowledgments}

We are grateful to Twitter for providing public post data, to Tumblr for mobile scrolling data, to Craig Silverman for the Emergent data, and to James Gleeson, Karen Church, Senaka Buthpitiya, Mitesh Patel, and Giovanni Ciampaglia for helpful discussions and assistance in data analysis. This work was supported in part by the James S. McDonnell Foundation (grant 220020274) and the National Science Foundation (award CCF-1101743). X.Q. thanks the NaN group in the Center for Complex Networks and Systems Research (cnets.indiana.edu) for the kind hospitality during her stay at the Indiana University School of Informatics and Computing. She was supported by grants from the National Natural Science Foundation of China (No.~90924030), the China Scholarship Council, the ``Shuguang'' Project of Shanghai Education Commission (No.~09SG38), and the Program of Social Development of Metropolis and Construction of Smart City (No.~085SHDX001). The funders had no role in study design, data collection and analysis, decision to publish, or preparation of the manuscript.

\subsection*{Author contributions}

A.F. and F.M. developed the research question. X.Q., D.F.M.O., A.F. and F.M. designed the model. X.Q. and D.F.M.O. conducted the simulations and the primary analyses. D.F.M.O., A.S. and F.M. collected and analyzed the empirical data. D.F.M.O., A.F. and F.M. wrote the manuscript. X.Q. and  A.S. edited the manuscript. 

\end{document}